\documentclass[preprint,12pt]{elsarticle}

\usepackage{amssymb}

\usepackage{graphicx}   

\journal{Physica A}

\begin{document}

\begin{frontmatter}



\title{Exact solution of a generalized version\\ of the Black-Scholes equation}


\author{L.-A. Cotfas\corref{cor1}\fnref{lac}}
\address[lac]{Faculty of Economic Cybernetics, Statistics and Informatics, Bucharest University of Economic Studies, 6 Piata Romana, 010374 Bucharest, Romania}
\ead{lcotfas@gmail.com}
\cortext[cor1]{Corresponding author}
\author{C. Delcea\fnref{lac}}
\ead{camelia.delcea@yahoo.com}
\author{N. Cotfas\fnref{nc}}
\address[nc]{University of Bucharest,  Physics Department, P.O. Box MG-11, 077125 Bucharest, Romania}
\ead{ncotfas@yahoo.com}

\begin{abstract}
We analyze a generalized version of the Black-Scholes equation depending on a parameter $a\!\in \!(-\infty ,0)$. It satisfies the martingale condition and coincides with the Black-Scholes equation in
the limit case $a\nearrow 0$. We show that the generalized equation is exactly solvable in terms of 
Hermite polynomials and numerically compare its solution with the solution of the Black-Scholes equation.
\end{abstract}

\begin{keyword}
econophysics\sep quantum finance  \sep Black-Scholes equation  \sep option pricing 
\MSC[2010] 91B80 \sep 91G80
\end{keyword}

\end{frontmatter}


\section{Introduction}
\label{introd}
The mathematical model based on the Black-Scholes equation
\begin{equation}\label{BS}
\frac {\partial C}{\partial t}=-\frac{\sigma ^2}{2}S^2\frac{\partial ^2C}{\partial S^2}-rS\frac{\partial C}{\partial S}+rC
\end{equation}
anticipates rather well the observed prices for options in the case of a strike price that is not too far from the current price of the underlying asset \cite{Ugur}. The price of an option at a moment of time $t$ depends on the current price $S$, the volatility $\sigma $, the risk-free interest rate $r$, the strike price $K$ and the maturity time $T$. In the case of an European option, the price is described by the solution $C(S,t)$ of equation 
(\ref{BS}) satisfying the condition
\begin{equation}\label{cond1}
C(S,T)=\left\{ 
\begin{array}{cll}
S\!-\!K & \mbox{if} & S\!\geq \!K\\[2mm]
0 & \mbox{if} & S\!<\!K
\end{array}\right.
\end{equation}
in the case of a call option, and
\begin{equation}\label{cond2}
C(S,T)=\left\{ 
\begin{array}{cll}
0 & \mbox{if} & S\!\geq \!K\\[2mm]
K\!-\!S & \mbox{if} & S\!<\!K
\end{array}\right.
\end{equation}
in the case of a put option. 
The alternative version of the equation (\ref{BS})
\begin{equation}
\label{BSS}
\frac {\partial \tilde C}{\partial t}=-\frac{\sigma ^2}{2}\frac{\partial ^2\tilde C}{\partial x^2}+\left(\frac{\sigma ^2}{2}-r\right)\frac{\partial \tilde C}{\partial x}+r\tilde C
\end{equation}
obtained by using the change of independent variable 
\begin{equation}
S={\rm e}^x
\end{equation}
allows one to use the formalism of quantum mechanics in option pricing \cite{Baaquie,Bagarello,LCotfas,Jana}.
In the new variable the conditions (\ref{cond1}) and (\ref{cond2}) become
\begin{equation}\label{cond1a}
\tilde C(x,T)=\left\{ 
\begin{array}{cll}
{\rm e}^x\!-\!K & \mbox{if} & x\!\geq \!\ln K\\[2mm]
0 & \mbox{if} & x\!<\!\ln K
\end{array}\right.
\end{equation}
and respectively,
\begin{equation}\label{cond2a}
\tilde C(x,T)=\left\{ 
\begin{array}{cll}
0 & \mbox{if} & x\!\geq \!\ln K\\[2mm]
K\!-\!{\rm e}^x & \mbox{if} & x\!<\!\ln K\, .
\end{array}\right.
\end{equation}

The more general version of (\ref{BSS}) depending on a function $V(x)$
\begin{equation}
\label{gBSS}
\frac {\partial \tilde C}{\partial t}=-\frac{\sigma ^2}{2}\frac{\partial ^2\tilde C}{\partial x^2}+\left(\frac{\sigma ^2}{2}-V(x)\right)\frac{\partial \tilde C}{\partial x}+V(x)\, \tilde C
\end{equation}
satisfies the martingale condition \cite{Baaquie} and hence can be used for studying processes in finance.
Our purpose is to investigate the particular case
\begin{equation}
V(x)=ax+r
\end{equation}
that is, the equation
\begin{equation}
\label{pBS}
\frac {\partial \tilde C}{\partial t}=-\frac{\sigma ^2}{2}\frac{\partial ^2\tilde C}{\partial x^2}+\left(\frac{\sigma ^2}{2}-ax\!-\!r\right)\frac{\partial \tilde C}{\partial x}+(ax\!+\!r)\tilde C
\end{equation}
where $a\!\in \!(-\infty ,0)$ is a parameter. The equation (\ref{pBS}) is exactly solvable in terms of Hermite polynomials and  coincides in the limit case $a\nearrow 0$  with the equation (\ref{BSS}) which corresponds to the standard Black-Scholes equation.

\section{A shifted oscillator}

Let $\alpha \!\in \! (-\infty ,0)$ and $\beta \!\in \! \mathbb{R}$ be two constants. By using the Hermite
polynomial
\begin{equation}
{\bf H}_n(s)=(-1)^n\ {\rm e}^{s^2}\ \frac{d^n}{ds^n}({\rm e}^{-s^2})
\end{equation}
we define for each  $n\!\in \!\{ 0,1,2,...\}$ the function
\begin{equation}
\label{psi}
\begin{array}{l}
\psi _n(x)=\frac{1}{\sqrt{n!\, 2^n}}\sqrt[4]{\frac{-\alpha }{2\pi }}\ {\rm e}^{\frac{\alpha }{4}x^2
+\frac{\beta }{2}x+\frac{\beta ^2}{4\alpha }}\ {\bf H}_n\left(\sqrt{\frac{-\alpha }{2}}x-\frac{\beta }{\sqrt{-2\alpha }} \right).
\end{array}
\end{equation}
If we denote
\begin{equation}
\begin{array}{l}
s=\sqrt{\frac{-\alpha }{2}}x-\frac{\beta }{\sqrt{-2\alpha }}
\end{array}
\end{equation}
then the previous relation can be written as
\begin{equation}
\begin{array}{l}
{\bf H}_n(s)=\sqrt{n!\, 2^n}\, \sqrt[4]{\frac{2\pi }{-\alpha }}\ {\rm e}^{\frac{1}{2}s^2}\ \psi _n\left(\sqrt{\frac{2}{-\alpha }}s-\frac{\beta }{\alpha }\right).
\end{array}
\end{equation}
By substituting this relation into the diferential equation
\begin{equation}
{\bf H}''_n(s)-2 s {\bf H}_n'(s)+2n {\bf H}_n(s)=0
\end{equation}
satisfied by the Hermite polynomial ${\bf H}_n$ we get the equality
\begin{equation}
\begin{array}{l}
-\psi _n''\left(\sqrt{\frac{2}{-\alpha }}s-\frac{\beta }{\alpha }\right)+\left(-\frac{\alpha }{2}s^2+\frac{\alpha }{2}+\alpha n\right)\ \psi _n\left(\sqrt{\frac{2}{-\alpha }}s-\frac{\beta }{\alpha }\right)=0
\end{array}
\end{equation}
which can be written in the form
\begin{equation}
\left(-\frac{\partial ^2}{\partial x^2}+\frac{(\alpha x\!+\!\beta)^2}{4}+\frac{\alpha }{2}\right)\psi _n=-\alpha n\, \psi _n.
\end{equation}
This means that $\psi _n$ is an eigenfunction of the shifted oscillator \cite{Cooper,NLCotfas,Jafarizadeh}.
\begin{equation}
H=-\frac{\partial ^2}{\partial x^2}+\frac{(\alpha x\!+\!\beta)^2}{4}+\frac{\alpha }{2}
\end{equation}
corresponding to the eigenvalue $\lambda _n\!=\!-\alpha n$, for any $n\!\in \!\{ 0,1,2,...\}$. Since
\begin{equation}
\int_{-\infty }^\infty {\rm e}^{-s^2}\ {\bf H}_n(s)\ {\bf H}_k(s)\ ds=\left\{
\begin{array}{cll}
n!\, 2^n\sqrt{\pi } & {\rm if} & n\!=\!k\\[2mm]
0  & {\rm if} & n\!\neq \!k
\end{array} \right.
\end{equation}
the system of functions $\{ \psi _n\}_{n=0,1,2,...}$ is orthonormal, that is,  
\begin{equation}
\int_{-\infty }^\infty \psi _n(x)\ \psi _k(x)\ dx=\left\{
\begin{array}{cll}
1 & {\rm if} & n\!=\!k\\[2mm]
0  & {\rm if} & n\!\neq \!k.
\end{array} \right.
\end{equation}
One can prove that it is complete in the space of square integrable functions.

\section{A generalized version of the Black-Scholes equation}

A straightforward generalization of the equation
\begin{equation}
\frac {\partial \tilde C}{\partial t}=-\frac{\sigma ^2}{2}S^2\frac{\partial ^2\tilde C}{\partial S^2}-rS\frac{\partial \tilde C}{\partial S}+r\tilde C
\end{equation}
is the equation
\begin{equation}
\frac {\partial \tilde C}{\partial t}=-\frac{\sigma ^2}{2}\frac{\partial ^2\tilde C}{\partial x^2}+\left(\frac{\sigma ^2}{2}-V(x)\right)\frac{\partial \tilde C}{\partial x}+V(x)\, \tilde C
\end{equation}
satisfying the martingale condition \cite{Baaquie}. It can be written in the form
\begin{equation}
\label{HgBSS}
\frac {\partial \tilde C}{\partial t}=H_V\tilde C
\end{equation}
by using the Hamiltonian
\begin{equation}
H_V=-\frac{\sigma ^2}{2}\frac{\partial ^2}{\partial x^2}+\left(\frac{\sigma ^2}{2}-V(x)\right)\frac{\partial }{\partial x}+V(x).
\end{equation}
The Hamiltonian $H_V$ is equivalent with the Hermitian Hamiltonian 
\begin{equation}
H_{\rm eff}=-\frac{\sigma ^2}{2}\frac{\partial ^2}{\partial x^2}+
\frac{1}{2}\frac{\partial V}{\partial x}+\frac{1}{2\sigma ^2}V^2+\frac{1}{2}V+\frac{\sigma ^2}{8}
\end{equation}
by the similarity transformation
\begin{equation}
H_{\rm eff}={\rm e}^{-u}H_V{\rm e}^u
\end{equation}
where
\begin{equation}
u=\frac{1}{2}x-\frac{1}{\sigma ^2}\int_0^xV(y)\, dy.
\end{equation}
In the particular case $V(x)\!=\!ax\!+\!r$ considered in this article
\begin{equation}
\begin{array}{l}
H_V=-\frac{\sigma ^2}{2}\frac{\partial ^2}{\partial x^2}+\left(\frac{\sigma ^2}{2}-ax-r\right)\frac{\partial }{\partial x}+ax+r\\[4mm]
H_{\rm eff}=\frac{\sigma ^2}{2}\left[ -\frac{\partial ^2}{\partial x^2}+\frac{1}{4}\left( \frac{2a}{\sigma ^2}x+\frac{2r}{\sigma ^2}+1 \right)^2+\frac{a}{\sigma ^2}\right]\\[4mm]
u=-\frac{a}{2\sigma ^2}x^2-\left(\frac{r}{\sigma ^2}-\frac{1}{2}\right)x.
\end{array}
\end{equation}
The Hamiltonian $H_{\rm eff}$ is up to the multiplicative factor $\frac{\sigma ^2}{2}$ the Hamiltonian of a shifted oscillator, namely
\begin{equation}
H_{\rm eff}=\frac{\sigma ^2}{2}H
\end{equation}
where
\begin{equation}
\begin{array}{l}
H=-\frac{\partial ^2}{\partial x^2}+\frac{(\alpha x\!+\!\beta)^2}{4}+\frac{\alpha }{2}
\end{array}
\end{equation}
with
\[
\begin{array}{l}
\alpha =\frac{2a}{\sigma ^2}\qquad \mbox{and}\qquad \beta=\frac{2r}{\sigma ^2}\!+\!1.
\end{array}
\]
Therefore, for each $n\in \{ 0,1,2,...\}$ the function
\begin{equation}
\begin{array}{l}
\psi _n(x)\!=\!\frac{1}{\sqrt{n!\, 2^n\sigma }}\sqrt[4]{\frac{-a }{\pi }}\ {\rm e}^{\frac{a}{2\sigma ^2}x^2+\left(\frac{r}{\sigma ^2}+\frac{1}{2}\right)x+\frac{\sigma ^2}{8a}\left(\frac{2r}{\sigma ^2}+1\right)^2}
{\bf H}_n\left( \frac{\sqrt{-a}}{\sigma }x-\frac{\sigma }{2\sqrt{-a}}\left(\frac{2r}{\sigma ^2}\!+\!1\right)\right)
\end{array}
\end{equation}
is an eigenfunction of $H_{\rm eff}$ corresponding to the eigenvalue $-an$
\begin{equation}
\begin{array}{l}
H_{\rm eff}\, \psi _n=-an\, \psi _n.
\end{array}
\end{equation}

The equation (\ref{HgBSS}) can be written as
\begin{equation}
\frac{\partial \tilde C}{\partial t}={\rm e}^uH_{\rm eff}{\rm e}^{-u}\, \tilde C
\end{equation}
or in the form
\begin{equation}
\frac{\partial }{\partial t}\,  {\rm e}^{-u} \tilde C=H_{\rm eff}\, {\rm e}^{-u} \tilde C .
\end{equation}
A function $\tilde C(x,t)$ is the solution of (\ref{HgBSS}) satisfying (\ref{cond1a})
if and only if
\begin{equation}
\psi (x,t)={\rm e}^{-u}\tilde C(x,t)
\end{equation}
that is, the function
\begin{equation}
\psi (x,t)={\rm e}^{\frac{a}{2\sigma ^2}x^2+\left(\frac{r}{\sigma ^2}-\frac{1}{2}\right)x}\tilde C(x,t)
\end{equation}
is the solution of the equation
\begin{equation}\label{eqpsi}
\frac{\partial  \psi }{\partial t}=H_{\rm eff}\, \psi 
\end{equation}
satisfying the condition
\begin{equation}\label{cond1b}
\psi (x,T)={\rm e}^{\frac{a}{2\sigma ^2}x^2+\left(\frac{r}{\sigma ^2}-\frac{1}{2}\right)x}\left\{ 
\begin{array}{cll}
{\rm e}^x\!-\!K & \mbox{if} & x\!\geq \!\ln K\\[2mm]
0 & \mbox{if} & x\!<\!\ln K\, .
\end{array}\right.
\end{equation}
But, the solution of (\ref{eqpsi}) satisfying (\ref{cond1b}) is
\begin{equation}
\psi (x,t)=\sum_{n=0}^\infty c_n\, {\rm e}^{-ant}\ \psi _n(x)
\end{equation}
with the coefficients $c_n$ determined from the relation
\begin{equation}
\sum_{n=0}^\infty c_n\, {\rm e}^{-anT}\ \psi _n(x)={\rm e}^{\frac{a}{2\sigma ^2}x^2+\left(\frac{r}{\sigma ^2}-\frac{1}{2}\right)x}\left\{ 
\begin{array}{cll}
{\rm e}^x\!-\!K & \mbox{if} & x\!\geq \!\ln K\\[2mm]
0 & \mbox{if} & x\!<\!\ln K\, 
\end{array}\right.
\end{equation}
namely,
\begin{equation}
c_n={\rm e}^{anT} \int_{\ln K}^\infty {\rm e}^{\frac{a}{2\sigma ^2}x^2+\left(\frac{r}{\sigma ^2}-\frac{1}{2}\right)x}\, ({\rm e}^x\!-\!K)\psi_n(x)\, dx.
\end{equation}
In the case of the call option, the  solution of the generalized Black-Scholes equation expressed in terms of Hermite polynomials is
\begin{equation}
\begin{array}{l}
C_a(S,t)\!=\!\frac{S}{\sqrt{\sigma }}\sqrt[4]{\frac{-a}{\pi }}{\rm e}^{\frac{\sigma ^2}{8a}\left(\frac{2r}{\sigma ^2}+1\right)^2}\!\sum\limits_{n=0}^\infty c_n \frac{{\rm e}^{-ant}}{\sqrt{n!\, 2^n}}\,  {\bf H}_n\!\left( \frac{\sqrt{-a}}{\sigma }\ln S\!-\!\frac{\sigma }{2\sqrt{-a}}\left(\frac{2r}{\sigma ^2}\!+\!1\right)\right).
\end{array}
\end{equation}
The case of a put option can be analyzed in a very similar way.

The Black-Scholes equation is  exactly solvable. By denoting
\begin{equation}
\Phi (\zeta )=\frac{1}{\sqrt{2\pi }}\int_{-\infty }^\zeta {\rm e}^{-\eta /2}d\eta =\frac{1}{2}\left( 1+{\rm erf}(x/\sqrt{2})\right)
\end{equation}
the formulas for the values of a European option can be written in the form
\begin{equation}
\begin{array}{l}
C_{\rm call}(S,t)=S\,  \Phi (d_1)-K\, {\rm e}^{-r(T-t)}\, \Phi (d_2)\\[2mm]
C_{\rm put}(S,t)=K\, {\rm e}^{-r (T-t)}\, \Phi (d_1)-S\,  \Phi (d_2)
\end{array}
\end{equation}
where \cite{Ugur}
\begin{equation}
\begin{array}{l}
d_1=\frac{ \ln (S/K)+\left(r+\frac{1}{2}\sigma ^2\right)(T-t)}{\sigma \sqrt{T-t}},\qquad 
d_2=\frac{ \ln(S/K)+\left(r -\frac{1}{2}\sigma ^2\right)(T-t)}{\sigma \sqrt{T-t}}.
\end{array}
\end{equation}
In Fig. 1 we present $C_a(S,t)$ (thick line) versus $C_{\rm call}(S,t)$ (thin line) and $C_{\rm call}(S,T)$
 (dashed line) for $t=3$ (left side), $t=4$ (right side) and $a=-0.03$ (first row), $a=-0.02$ (second row),
  $a=-0.01$ (third row) by choosing the following values of the parameters:  $\sigma \!=\!0.25$, $r\!=\!0.03$, $K\!=\!3$ and $T\!=\!5$.
  
\begin{figure}[t]
\centering
\includegraphics[scale=0.7]{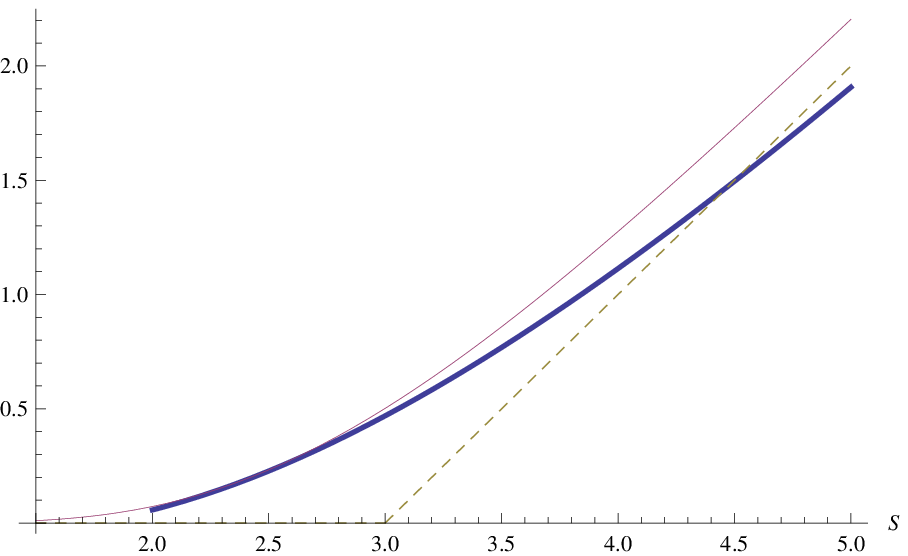}\qquad 
\includegraphics[scale=0.7]{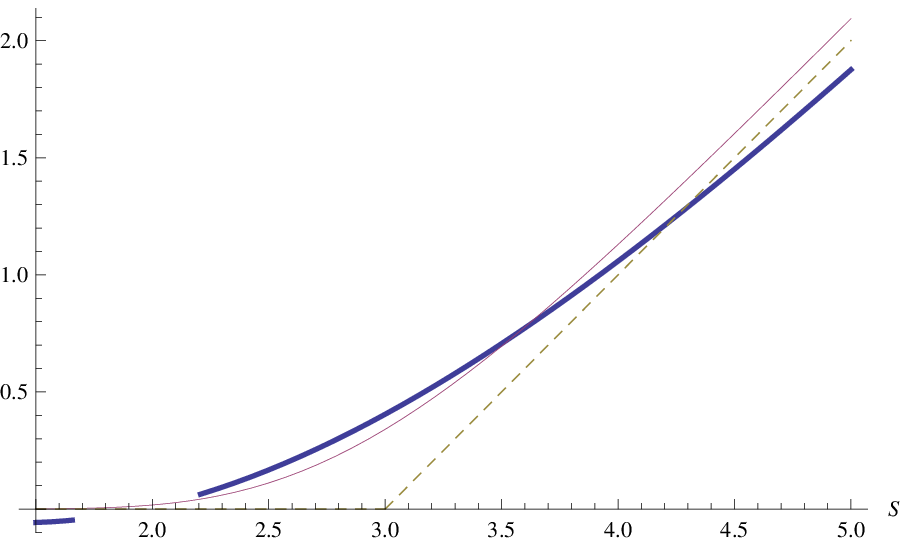}\\[3mm]
\includegraphics[scale=0.7]{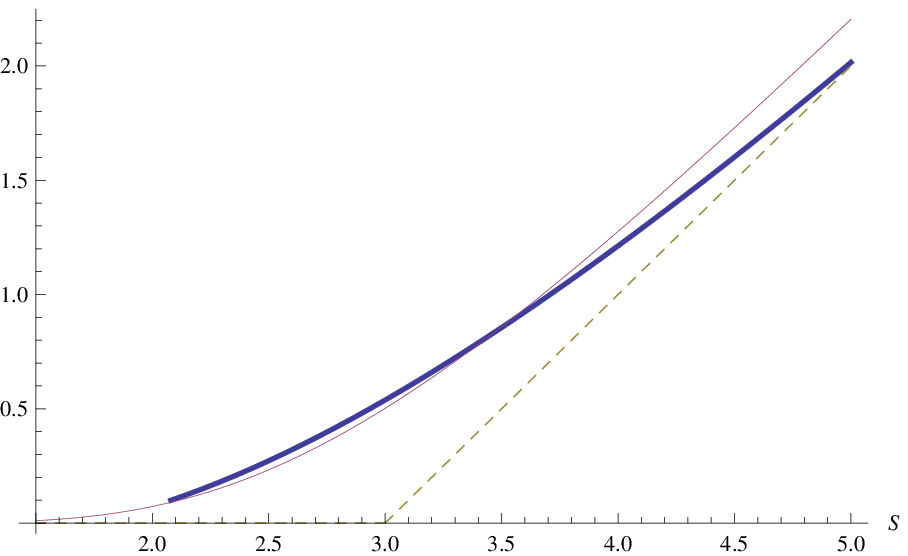}\qquad 
\includegraphics[scale=0.7]{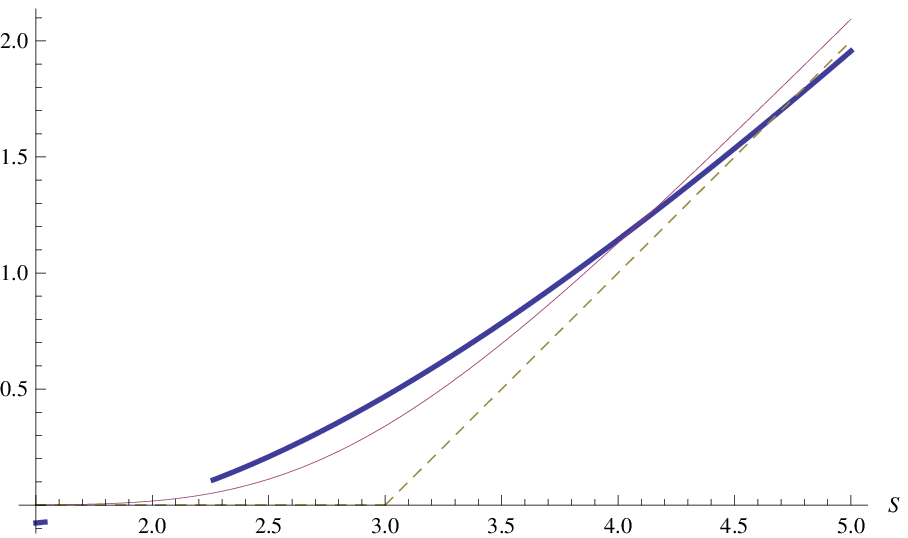}\\[3mm]
\includegraphics[scale=0.7]{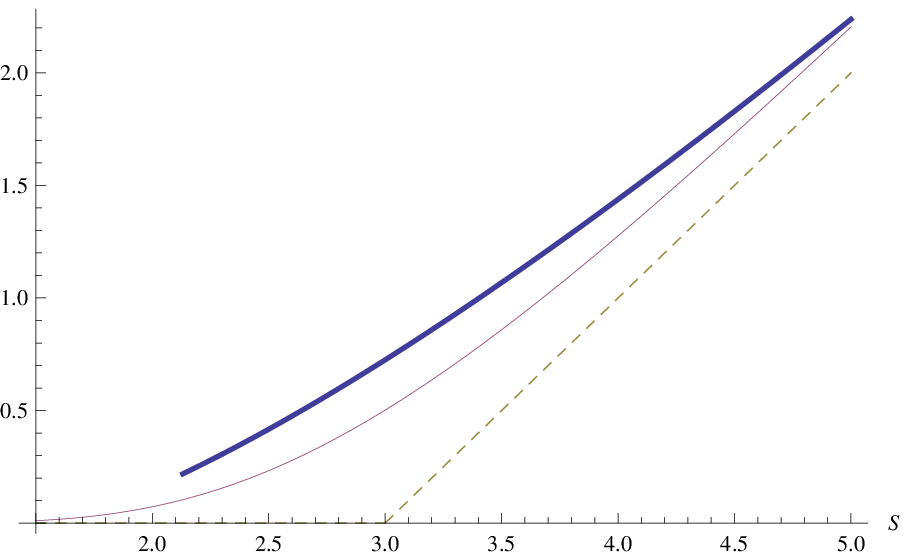}\qquad 
\includegraphics[scale=0.7]{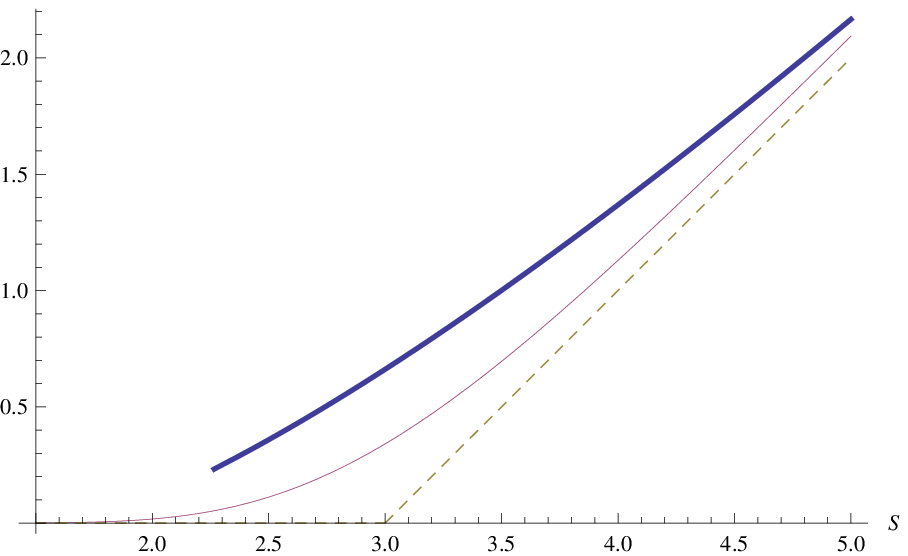}
\caption{The solution $C_a(S,t)$ (thick line) of the generalized Black-Scholes equation versus the solution $C_{\rm call}(S,t)$ (thin line) of the Black-Scholes equation and the payoff function $C_{\rm call}(S,T)$
 (dashed line) for $t=3$ (left hand side), $t=4$ (right hand side) and $a=-0.03$ (first row), $a=-0.02$ (second row),
  $a=-0.01$ (third row). }
\end{figure} 

\section{Concluding remarks}

In quantum mechanics as well as in econophysics, exact solutions are known only in a small 
number of particular cases and, generally, they play an important role. 
The generalized version of the Black-Scholes equation investigated in this article:

- is exactly solvable for any $a\!\in \!(-\infty ,0)$,

- satisfies the martingale condition  for any $a\!\in \!(-\infty ,0)$,

- coincides with the Black-Scholes equation in the limit case $a\nearrow 0$.\\
In practice, there are some deviations of prices from those described by the solution of 
the Black-Scholes equation. We think that the solution of the generalized Black-Scholes 
equation might describe some observed prices, and the parameter $a$ might have a certain financial meaning.








\end{document}